\pdfoutput=1
\documentclass[12pt,amsmath,amssymb,longbibliography,nofootinbib,showkeys,eprint]{revtex4-2}
\usepackage[activate={true,nocompatibility},final,tracking=true,kerning=true,spacing=true]{microtype}
\usepackage{epigraph}
\usepackage{graphicx}
\usepackage{siunitx}
\usepackage{seqsplit}
\usepackage{array}
\usepackage[T1,T2A]{fontenc}
\usepackage[utf8]{inputenc}
\usepackage{caption}
\usepackage{subcaption}
\usepackage{epstopdf}
\usepackage{numprint}
\usepackage{amsmath}
\usepackage{amssymb}
\usepackage{circuitikz}
\usepackage[colorlinks = false]{hyperref}

\newcommand\CapBlueLink{\href{http://datasheets.avx.com/AVX-SCC-LE.pdf}{AVX-SCCS20B505PRBLE}}
\newcommand\CapGreenLink{\href{https://www.eaton.com/content/dam/eaton/products/electronic-components/resources/data-sheet/eaton-hv-supercapacitors-cylindrical-cells-data-sheet.pdf}{Eaton-HV1020-2R7505-R}}
\newcommand\CapBlackLink{\href{https://ru.mouser.com/datasheet/2/88/DCN_series-553005.pdf}{IC-505DCN2R7Q}}
\newcommand\CapWhiteLink{\href{https://ru.mouser.com/datasheet/2/257/CE_2017_Datasheet_2_7V5F_3001974_EN_1-1274199.pdf}{Nesscap-ESHSR-0005C0-002R7}}

\begin{document}
\title{On The Inverse Relaxation Approach To Supercapacitors Characterization}

\author{Mikhail Evgenievich {Kompan}}
\email{kompan@mail.ioffe.ru}
\author{Vladislav Gennadievich {Malyshkin}} 
\email{malyshki@ton.ioffe.ru}
\address{Ioffe Institute, St. Petersburg, Russia, 194021}

\date{July 7, 2019}

\begin{abstract}
\begin{verbatim}
$Id: inverserelaxation.tex,v 1.217 2020/12/02 17:50:51 mal Exp $
\end{verbatim}
A novel inverse relaxation technique for supercapacitor
characterization is developed, modeled numerically,
and experimentally tested on a number of commercial supercapacitors.
It consists in
shorting a supercapacitor
for a short time $\tau$,
then switching to the open circuit regime and measuring an initial
rebound and long--time relaxation.
The results obtained are:
the ratio of ``easy'' and ``hard''
to access capacitance
and the dependence $C(\tau)$,
that determines
what the capacitance the system 
responds at time--scale $\tau$;
it can be viewed as an alternative to used by some manufacturers
approach to characterize a supercapacitor
by fixed capacitance and time--scale dependent internal resistance.
Among the advantages of proposed technique
is that it does not require a source of fixed current,
what simplifies the setup and allows a high discharge current
regime.
The approach
can be used as a replacement of low--frequency
impedance measurements and the ones of IEC 62391 type, it
can be effectively applied
to characterization
of supercapacitors 
and other relaxation type systems with porous internal structure.
The technique can be completely automated
by a microcontroller
to measure, analyze, and output
the results.
\end{abstract}

\maketitle

\newpage
\begin{flushright}
  {\small Dedicated to the memory of S.L. Kulakov}
\end{flushright}

\section{\label{intro}Introduction}
A distributed internal RC structure is manifested
in electrical measurements of supercapacitors.
The distribution is caused by hierarchical
porous structure of electrodes.
The two most commonly used technologies for manufacturing
carbon structures for supercapacitor electrodes are
{Carbide--derived carbon (CDC)}
and {Activated carbon}.
CDC
 materials are derived from carbide precursors\cite{oschatz2017carbide}.
  An initial crystal structure of the carbide is the primary factor affecting
  CDC porosity.
Activated carbon
  is typically derived from a charcoal or biochar\cite{abioye2015recent}.
  It's structure is inherited from the starting material and
  has a surface area in excess of $2,000m^2/g$ \cite{mangun2001surface}.
See \cite{borenstein2017carbon} for a review of
carbon materials used in supercapacitor electrodes.
All the technologies used for supercapacitor manufacturing
lead to a complex, ``self--assembled'' type of internal structure.
In applications
the most interesting is not the internal structure of a device per se,
but it's manifestation in the electric properties.

While Li--ion systems are the most effective
in energy storage applications\cite{du2003comparative},
supercapacitors are the most effective
in high--power applications\cite{burke2011power}.
For Li--ion batteries
the two characteristics
are typically provided by manufacturers:
specific energy and specific power.
For supercapacitors the other two characteristics
are typically provided by manufacturers:
capacitance and internal resistance.
Standard methods of characterization
create a substantial uncertainty,
because a supercapacitor's characteristics change during
the discharge process.

The techniques currently used for characterization of
a supercapacitor's electric properties
can be classified as:
\begin{itemize}
\item
Low--current impedance AC spectroscopy
is a frequency domain technique.
High frequency range $10^{-3} \div 10^6$Hz
allows an information of porous structures
to be obtained. However  a low-current measurement
regime, interpretation difficulties, and equipment
complexity limit the technique applicability.
\item
Cyclic Voltammetry is a time--domain technique,
where the voltage is swept between lower and upper limits at
a fixed scan rate.
The voltage scan rate is the slope of the $U(t)$;
current evolution is measured as a function of the voltage.
This technique is quite common for electrochemical materials
study, it is less convenient for supercapacitors characterization,
where a high current regime is often required.
\item Constant current charge/discharge regime is the most used
time--domain technique for supercapacitors characterization.
A measurement starts by switching to a given constant charging current.
The initial instantaneous voltage jump
determines  the capacitor series resistance.
The current is switched off at time when the capacitor has
reached the maximal voltage,
the voltage  instantaneously drops due current interruption
via the series resistance. Then similar processes take place
in the discharge regime.
The value
of the current is defined in \cite{IEC62391} \texttt{IEC 62391-\{1,2\}}
standards.
The value of the internal resistance is determined from the potential jumps.
The value of the capacitance is determined from
the time necessary to charge/discharge the capacitor at given current.
\end{itemize}
Multiple extensions of the discharge techniques\cite{cheng2009assessments,yang2020comparative,zhang2015supercapacitors} have been recently proposed,
the most noticeable are: total charge measurement
and the difference of supercapacitors
characteristics in 
constant power and constant current discharge regimes\cite{burke2011power}.

In this paper a novel technique of
supercapacitor characterization
 is developed.
The technique has all the measurements
performed in time--domain, possibly at high current.
The results obtained are:
1. the ratio $\eta$ of ``easy'' and ``hard''
to access capacitance
and 2. the dependence $C(\tau)$,
that determines
what the capacitance the system 
responds at time--scale $\tau$,
varied in at least three orders range.
The technique was microcontroller--automated
to measure, analyze, and output
the results;
the measurement of $\eta$ does not require total change measurement,
but the measurement of $C(\tau)$ does
require one, it is implemented using
a microcontroller with ADC.
There are two distinguished features of the technique:
no fixed--current source requirement and the measurements are performed
at various time--scales $\tau$, what introduces a parameter $\tau$ similar to
inverse frequency in impedance spectroscopy technique.
The approach can be effectively used as a replacement of low--frequency
impedance measurements
to determine the internal resistance and capacitance.

\section{\label{TheModel}Inverse Relaxation Model}

\begin{figure}[t]
\begin{flushleft}
\begin{circuitikz}[american voltages, european resistors]
  \def\rcchain(#1,#2){
  \def\lastcnum{9}
  \def\xinit{#1}
  \def\yinit{#2}
  \foreach \cnum in {1,3,5,7,9} {
    \ifnum \lastcnum=\cnum {
       \draw (\xinit+\cnum,\yinit) [dashed] to[C] ++(0,-2)  to [R] ++(2,0);
       \draw (\xinit+\cnum,\yinit) [dashed] to [R] +(2,0);
    } \else {
      \draw (\xinit+\cnum,\yinit) to[C] ++(0,-2)  to [R] ++(2,0);
      \draw (\xinit+\cnum,\yinit) to [R] +(2,0);
    }\fi       
    \draw (\xinit+\cnum,\yinit)  [dashed] to [/tikz/circuitikz/bipoles/length=20pt,R] ++(1.3,1.2);
    \draw (\xinit+\cnum,\yinit)  [dashed] to ++(-0.4,0.5);
    \draw (\xinit+\cnum,\yinit-2)  [dashed] to [/tikz/circuitikz/bipoles/length=20pt,R] ++(1.1,1.2);
     \draw (\xinit+\cnum,\yinit-2)  [dashed] to ++(-0.4,0.5);
  }
  \draw (\xinit+1,\yinit) to [R] ++(-2,0);
  \draw (\xinit+1,\yinit-2) to [R] ++(-2,0);
  \draw (\xinit-1,\yinit)  [dashed] to [/tikz/circuitikz/bipoles/length=20pt,R] ++(1.3,1.2);
  \draw (\xinit-1,\yinit)  [dashed] to ++(-0.4,0.5);
  \draw (\xinit-1,\yinit-2)  [dashed] to [/tikz/circuitikz/bipoles/length=20pt,R] ++(1.1,1.2);
  \draw (\xinit-1,\yinit-2)  [dashed] to ++(-0.4,0.5);
}

\draw (0,2) to [open, v=SC] (0,0) ;
\rcchain(2,2)
\draw (0,2) to[short, *-] (1,2);
\draw (0,0) to[short, *-] (1,0);
\end{circuitikz} \\
\bigskip
\begin{circuitikz}[american voltages, european resistors]
  \draw (0,2) node[anchor=east] {} to[short, *-] (0,2)
  to[R=$R_1$] (2,2)
  to[R=$R_2$] (4,2) ;
  \draw (4,2) [dashed] to [R=$R_3$] (6,2);
  \draw (6,2) [dashed]  to  (7,2);
  \draw (0,0) node[anchor=east] {} to[short, *-] (6,0);
  \draw (6,0) [dashed] to (7,0);
  \draw (2,2) to[C=$C_1$] (2,0);
  \draw (4,2) to[C=$C_2$] (4,0);
  \draw (6,2) [dashed] to[C=$C_3$] (6,0);
  \draw (0,2) to [open, v=SC] (0,0);
\end{circuitikz}
\end{flushleft} 

\bigskip

\begin{circuitikz}[scale=1,line width=1pt, european resistors]
\makeatletter\@input{regswitchforoldcircuitikz.sty}\makeatother

  \draw (0,4) node[anchor=east] {$U(t)$} to[short, *-] ++(3,0)
  to[nos=$\mathrm{Charge}$] ++(2,0)
  to[short, -*] ++(0,0) 
  to node[anchor=west] {$U_0$} ++(0,0) ;
  \draw (3,4)
  to[nos=$\mathrm{Short}$] ++(0,-2)
  to [R=$R_s$] ++(0,-1.5)
  to[short, -] ++(0,-0.5) ;
  \draw (0,0) to[short, *-*] ++(5,0);
  \draw (2.5,2.2)  node[anchor=east] {$U^*(t)$}
  to[short, *-] ++(0.5,0);
  \draw (1,0) to[C={SC}] ++(0,3)
   to[short, -] ++(0,1);
\end{circuitikz}
\qquad
  \includegraphics[width=7cm,trim=0 1cm 0 0]{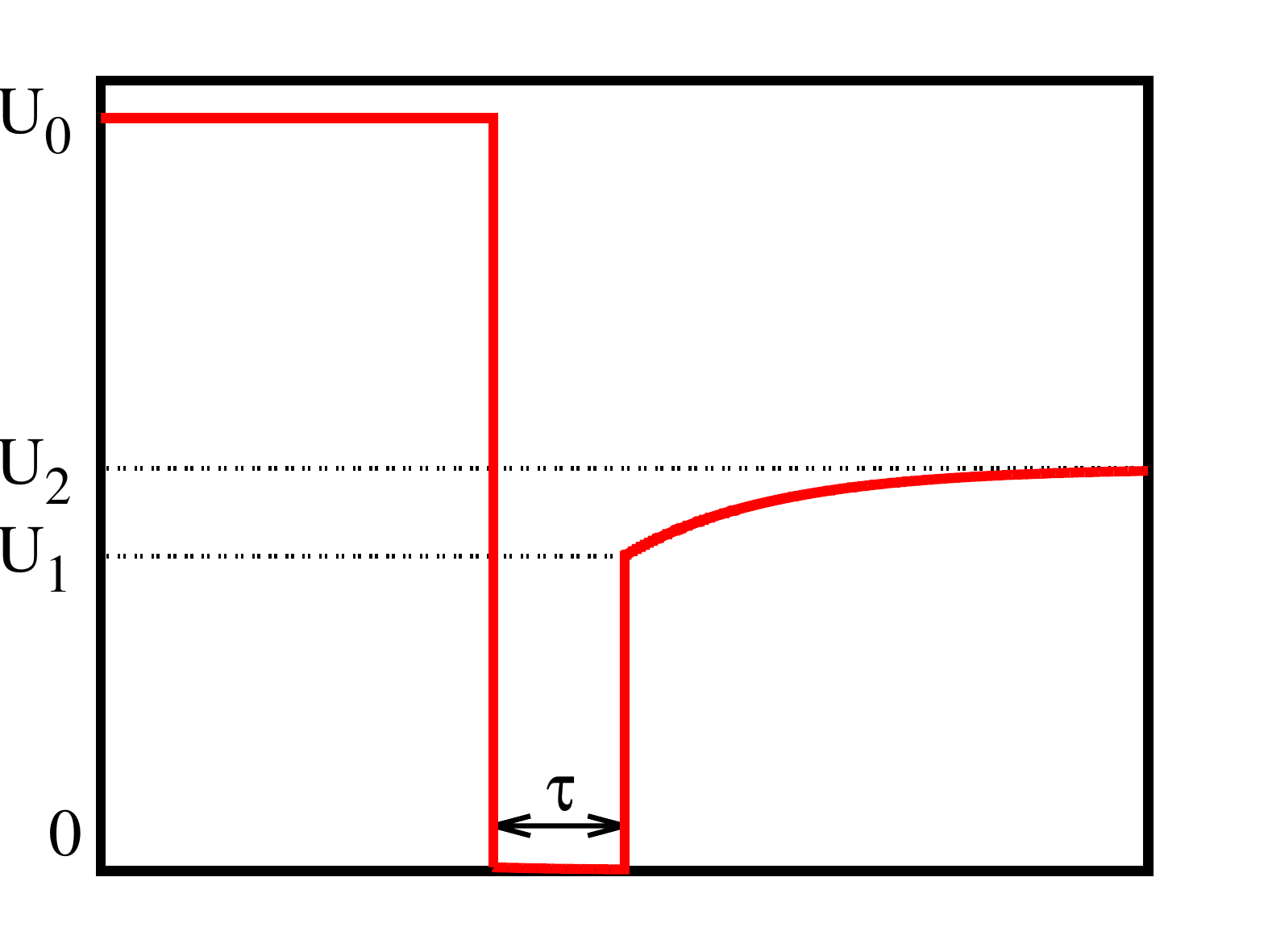}  
  \caption{\label{RCscheme}\label{Utshorting}
   Supercapacitor (SC) equivalent circuit
   corresponding to a hierarchical structure and the simplistic two--$RC$ model.
   Below are the measurement circuit and
  a general form of $U(t)$ dependence
  in inverse relaxation measurement technique.
1. Initial shorting for $\tau\ll RC$,
    the total charge passed is $Q$; the value of $R_s$ is very small.
    2. Immediate rise from $U_s\approx 0$ to $U_1$.
    3. In the open circuit regime
    a slow final rise from $U_1$ to $U_2$ due to internal charge
    redistribution.
  }
\end{figure}

Supercapacitor
equivalent circuit
has
multiple internal $RC$s, what corresponds
to it's
hierarchical internal structure\cite{yoo2016fast,borenstein2017carbon,kompan2019reverse,il2020modeling}. The dynamics of such a system is rather complex,
it is exhibited, for example, in multi--exponent
evolution of $U(t)$ relaxation (see experimental Fig. \ref{experimentatime} below) and in deviation from a rectangle in a cyclic
voltammetry (CV) plot.
For an application in electronics
the most convenient characteristic is:
how a supercapacitor behaves at a given time--scale $\tau$,
how much $Q$ it can be charged/discharged during time--interval $\tau$.
In terms of electric properties
supercapacitor's electrodes internal porous structure
can be considered as electric capacitance of
two kind:
``easy'' (accessible at low $\tau$)
and ``hard'' (accessible only at high $\tau$), Fig. \ref{RCscheme}.
Actual distribution of the internal $RC$ can be of various forms,
because the internal structure manifests itself in the distribution of $RC$.
This approach  is more objective compared to
impedance technique (which is a low amplitude technique),
since it characterizes the discharge as a whole.
When a supercapacitor is in the stationary state, all the capacitors
in Fig. \ref{RCscheme} model have equal potential,
it is equal to the one on the electrodes,
and there is no internal current. When a supercapacitor is in a
non--stationary state then internal charge redistribution takes place,
it can be directly observed through the dynamics of electrodes potential.

\begin{figure}[t]
\includegraphics[width=8cm]{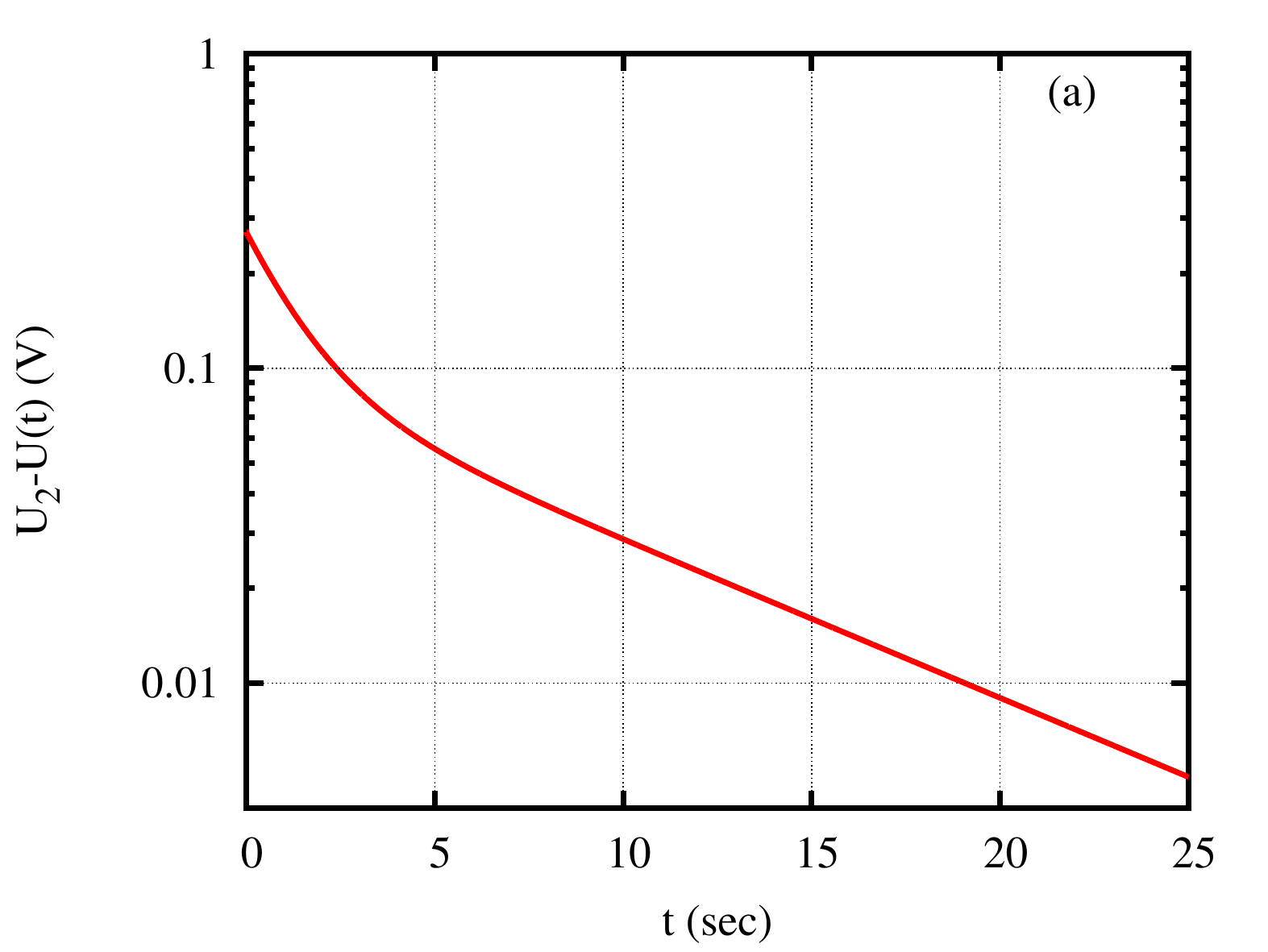}
\includegraphics[width=8cm]{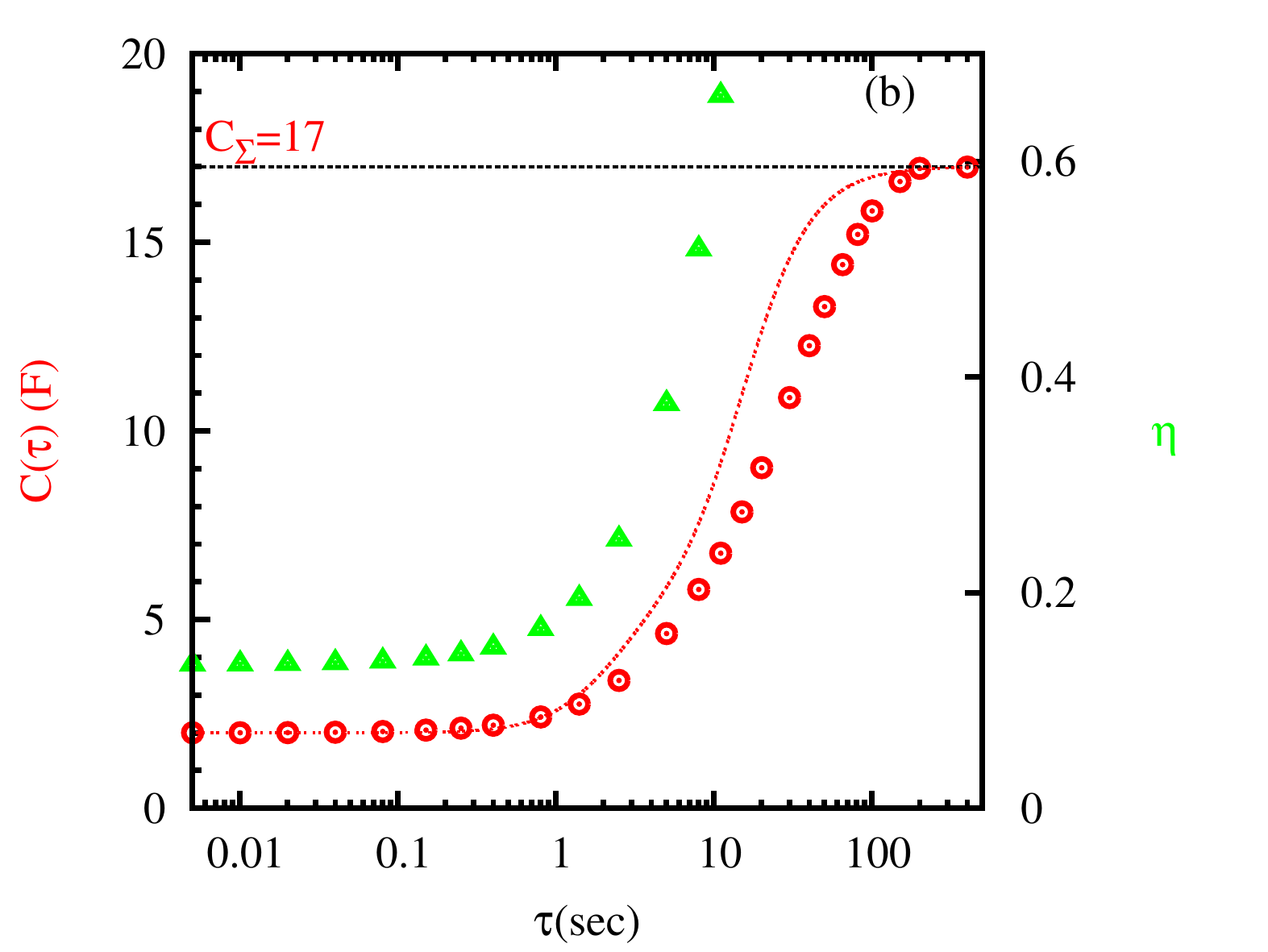}
\caption{
\label{threeRC}
A three--$RC$ system:
  $R_1=1\Omega$, $C_1=2F$, $R_2=1\Omega$, $C_2=5F$, $R_3=2\Omega$, $C_3=10F$.
  (a): $U_2-U(t)$ evolution;
  a deviation from a linear dependence ($U_2-U(t)$ is in log scale)
  is clearly observed.
  (b): $C(\tau)$ (circles, left axis) and $\eta$ (triangles, right axis).
  Dashed line corresponds to $C_{impedance}(\tau)$ from (\ref{Cimpedancetau})
  with the impedance $Z(1/\tau)$ for this three--$RC$ system.
  The value of $C_{\Sigma}$ from (\ref{CtauSum}) is a constant.
}
\end{figure}

Consider a measurement technique:
the system is charged to some
initial potential $U_0$, then it is short--circuited for a short interval time (lower
than the supercapacitor's internal $RC$) to create a non--stationary state,
after that
it is switched to the open circuit regime and $U(t)$ is recorded
to observe internal relaxation.
The $U(t)$ dependence is:
\begin{itemize}
\item First, from the initial potential $U_0$
  to almost zero (shorting
  to create a non--stationary state).
  Instead of shorting, a connection to a
  low--resistance circuit
  (we denote it $R_s$, a typical
  value is about $10-50m\Omega$)
can be used, Fig. \ref{Utshorting}, in this
  case the potential is non--zero,
 the potential at the
  moment right before switching to the open circuit regime
  is denoted
  as $U_s$.
\item Then, after switching to the open circuit regime,
the potential jumps from $U_s$ to $U_1$. There is a similar
current-‐interruption technique used
in fuel cell measurements \cite{larminie2003fuel}, page 64,
the immediate rise voltage $V=IR_i$
is an equivalent of $U_1-U_s$.
\item A slow final rise from $U_1$ to $U_2$, Fig. \ref{Utshorting}.
The $U(t)$ relaxation from $U_1$ to $U_2$
may be of a single of multiple exponent
type, this depends on the supercapacitor's internal structure.
For two--$RC$ model a pure linear dependence is observed,
For three--$RC$ supercapacitor model there are two exponents in $U(t)$ evolution,
one can clearly observe a deviation from a linear dependence in Fig. \ref{threeRC}a below.
\item While the $U(t)$ measurement can be performed using
traditional equipment a progress in microcontrollers
(e.g. the \href{https://www.st.com/en/microcontrollers-microprocessors/stm32-32-bit-arm-cortex-mcus.html}{STM32F103C8T6 ARM} which costs below \$5 and has a 72Mhz CPU with 12--bit analog--to--digital converter (ADC))
allows the data to be easily recorded and stored.
A microcontroller allows
the total charge
passed on the shorting stage
to be calculated by direct integration:
\begin{align}
Q(\tau)&=\int\limits_0^{\tau} I dt\approx\sum\limits_{k} \frac{U(t_k)}{R_s}(t_k-t_{k-1})
\label{Qdc}
\end{align}
\end{itemize}
``Right rectangle'' integration rule is used to simplify
microcontroller implementation, it is more that adequate for
a typical sampling frequency $10^5/sec$.

Before we consider a more realistic model,
let us demonstrate how
the ratio of easy and hard to access capacitance
can be found with  the inverse relaxation technique
for
a \textsl{two}--$RC$ model.
In this case the separation on ``easy'' and ``hard'' to access capacitance is trivial:
$C_1$ is easy to access, $C_2$ is hard to access.
In two--$RC$ model an internal charge redistribution between
$C_1$ and $C_2$ is:
\begin{eqnarray}
  \Delta Q_{C_1}&=&-\Delta Q_{C_2} \label{chargedisbalance}\\
  C_1\cdot(U_2-U_1)&=&C_2\cdot(U_0-U_2) \\
  \eta=\frac{C_1}{C_2}&=&\frac{U_0-U_2}{U_2-U_1} \label{Cratio}
\end{eqnarray}

\begin{figure}
 \includegraphics[width=8cm]{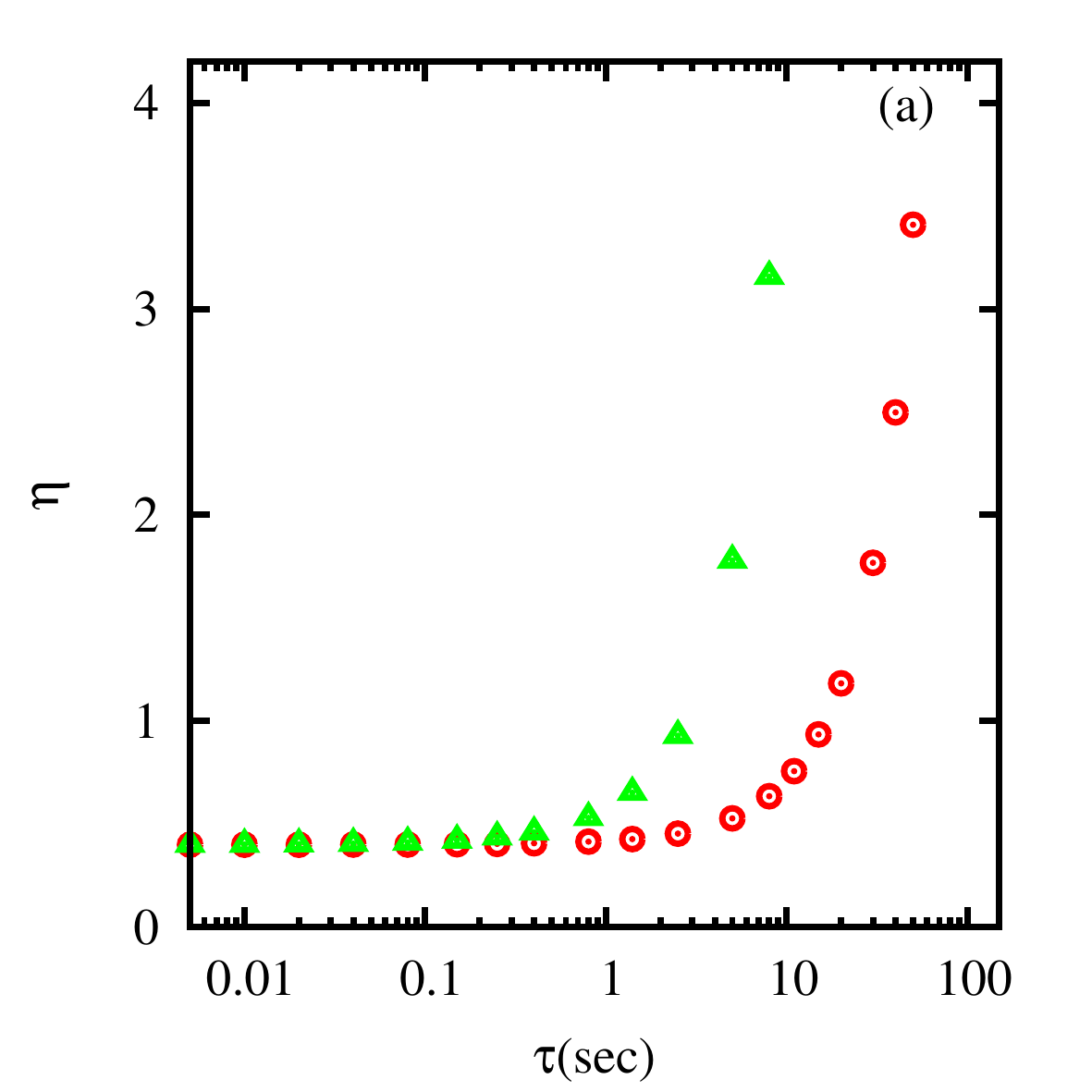}
 \includegraphics[width=8cm]{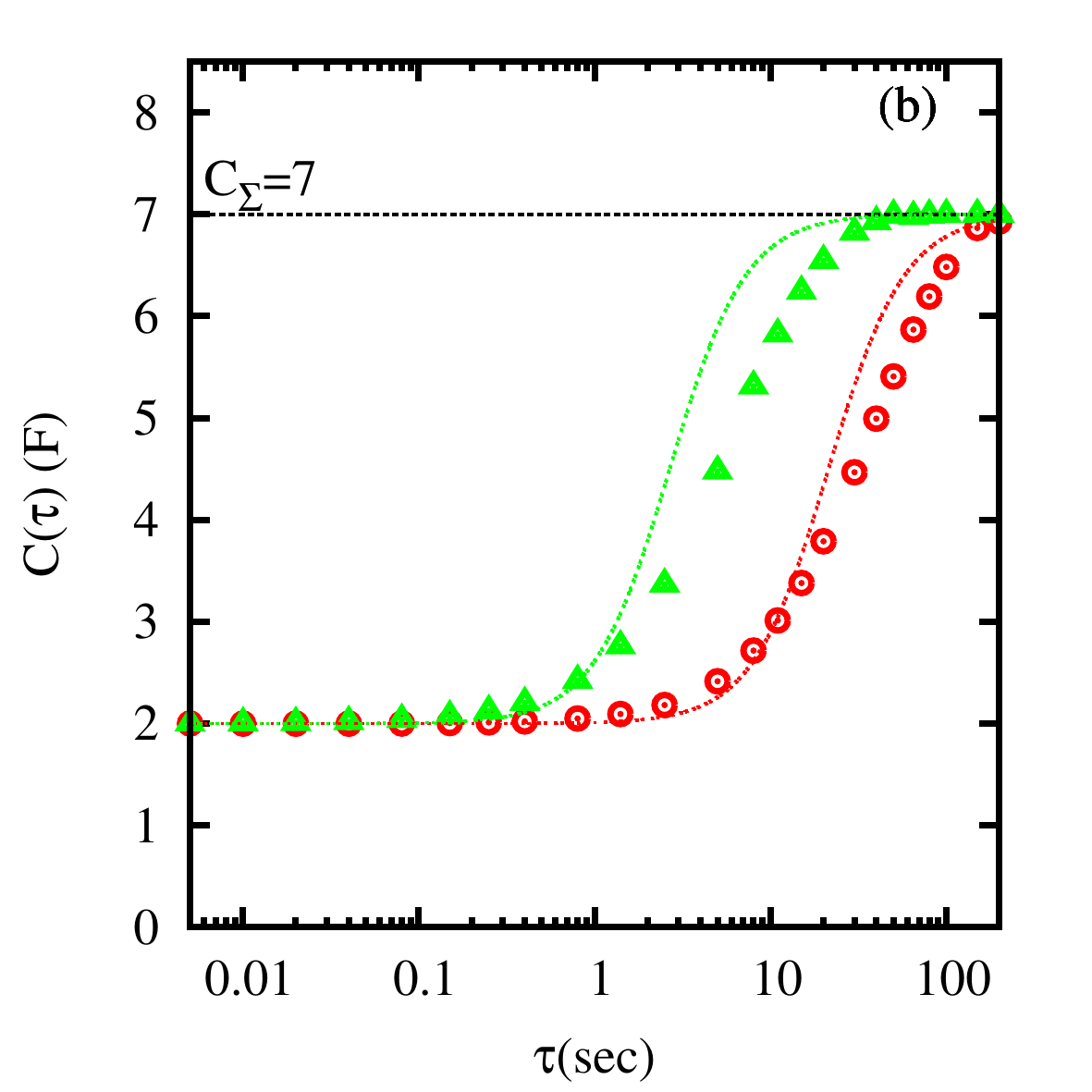}
\caption{\label{twoRC}
Two--$RC$ models:
   $R_1=1\Omega$, $C_1=2F$, $R_2=8\Omega$, $C_2=5F$ (circles) and
    $R_1=1\Omega$, $C_1=2F$, $R_2=1\Omega$, $C_2=5F$ (triangles).
The models are chosen to have identical
internal resistance $R_1$,
$\eta(\tau=0)={C_1}/{C_2}=0.4$, $C(\tau=0)=C_1=2F$, and $C(\tau=\infty)=C_1+C_2=7F$,
but eight times different inverse relaxation time $R_2C_1 C_2/(C_1+C_2)$.
  The  charts present the dependence  of (a):  $\eta=(U_0-U_2)/(U_2-U_1)$ on shorting time
    $\tau$, and (b): $C(\tau)=Q/(U_0-U_1)$ on shorting time
    $\tau$.   Dashed lines correspond to $C_{impedance}(\tau)$ from (\ref{Cimpedancetau}).
    The value of $C_{\Sigma}$ from (\ref{CtauSum}) is a constant.
}    
\end{figure}

Important, that the ratio (\ref{Cratio}) of ``easy''and  ``hard''
capacitance does not depend
on shorting time and on specific values of $R_1$ and $R_2$.
In two capacitors model
the values $U_0$, $U_1$, and $U_2$
can be obtained analytically,
but we are going to present a numerical solution with the goal
to study a more complex model later on.
In Fig. \ref{twoRC}a the
dependence of $\eta$ on $\tau$ is presented for
two different
two--$RC$ capacitor models with the same $\eta$. One can clearly see
that
the ratio (\ref{Cratio}) does not differ from the exact
value $C_1/C_2=0.4$ when 
shorting time $\tau$
changes in two orders of magnitude range.
A deviation from the constant 
arises only when
shorting time $\tau$ becomes comparable
to the supercapacitor's internal $RC$.
For a small
$\tau$
charge redistribution inside a supercapacitor
leads to $\eta$ independence on $\tau$ in a wide interval.
When one starts to increase the $\tau$ --- an initial
charge redistribution
becomes more  prolonged and the
deviation of $\eta$ from a constant
can be observed.
The independence of $\eta$ on $\tau$ allows us to consider
the ratio $\eta(\tau\to 0)$ as an immanent characteristic
of the system,
it is a characteristic
that separates easy  (accessible at low $\tau$)
and hard (accessible only at high $\tau$) to access capacitance.

The $\eta$ is obtained only from
the measurement of the potential (requires no charge measurement),
and, while useful for structural
characterization, lacks the information about absolute values.
To obtain these the total charge $Q$ passed on shorting stage is required,
this requires 
a microcontroller to calculate (\ref{Qdc}).
Once the $Q$ is obtained, absolute values of $R_1$ and $C_1$ are:
\begin{align}
C_1&\approx \frac{Q}{U_0-U_1} \label{Ci1measurement} \\
R_1&=\left[\frac{U_1}{U_s}-1\right]R_s \label{Ri}
\end{align}
The (\ref{Ci1measurement}) follows from $\tau\ll RC$;
the (\ref{Ri}) follows immediately from current balance: $U_s/R_s=\left[U_1-U_s\right]/R_1$.
Now consider an increase of $\tau$ to the values above $RC$. Then $R_1$
(\ref{Ri}) stays the same, it does not depend on $\tau$ at all.
The $C_1$ (\ref{Ci1measurement}) increases with $\tau$, in the limit of large $\tau$
it becomes equal to total capacitance $C_{\Sigma}=C_1+C_2+\dots$.
Introduce 
\begin{align}
C(\tau)&= \frac{Q(\tau)}{U_0-U_1(\tau)} \label{Ctau}
\end{align}
that determines what the capacitance the system responds at time--scale $\tau$.
A large $\tau$ corresponds to full discharge,
a small $\tau$ corresponds only to a
discharge of some porous branches (partial discharge).
In the  Fig. \ref{twoRC}b the dependence of $C(\tau)$ is presented
for two--$RC$ models.
As expected $C(\tau=0)=C_1$ and $C(\tau=\infty)=C_1+C_2+\dots$.
A similar to $C(\tau)$ concept can be conceived from impedance
consideration. For two--$RC$ model the impedance $Z(\omega)$ is:
\begin{align}
  Z_2(\omega)&=R_2+\frac{1}{j\omega C_2} \nonumber \\
  Z(\omega)&= R_1+ \frac{Z_2(\omega)}{1+Z_2(\omega)j\omega C_1}
  \label{zw2rc}
\end{align}
Then, formally considering a ``capacitance'' as inverse proportional
to imaginary part of $\omega Z(\omega)$ and time--scale as $\tau=1/\omega$
one can introduce:
\begin{align}
C_{impedance}(\tau)&= \frac{\tau}{- \mathrm{Im}(Z(1/\tau))}
\label{Cimpedancetau}
\end{align}
This definition treats impedance reactive component
as caused by some ``effective capacitance'', this is reasonable
while there is no inductances\cite{kompan2013inductive} in the system.
The $C_{impedance}(\tau)$
behaves similarly to $C(\tau)$,
e.g. it has the same asymptotic
$C_{impedance}(\tau=0)=C_1$ and $C_{impedance}(\tau=\infty)=C_1+C_2+\dots$.
It is presented in Fig. \ref{twoRC}b
to compare with $C(\tau)$.
One can see a similar
dependence.
The $C_{impedance}(\tau)$ from  (\ref{Cimpedancetau}) 
can be used for an arbitrary distributed $RC$ system,
not only for the two--$RC$ impedance $Z(\omega)$ from (\ref{zw2rc}), see
Fig. \ref{threeRC}b below for a three--$RC$ system example.
The major advantage of the $C(\tau)$ over the $C_{impedance}(\tau)$
is that it can be measured purely in time--domain.

If the inverse relaxation potential $U_2$ is used in (\ref{Ctau})
instead of the $U_1$ obtain the total capacitance
\begin{align}
C_{\Sigma}&= \frac{Q(\tau)}{U_0-U_2(\tau)} \label{CtauSum}
\end{align}
which is independent on the value of shorting time $\tau$,
but it is often difficult to measure $U_2$ experimentally at low  $\tau$
because of charge leaks. There is no such a difficulty to
measure the 
$C(\tau)$ from (\ref{Ctau}); for this reason it is
convenient to express  
the $\eta(\tau)$ from (\ref{Cratio}) via
the $C(\tau)$:
\begin{align}
C(\tau)&=C_{\Sigma}\frac{U_0-U_2}{U_0-U_1}=C_{\Sigma}\frac{\eta(\tau)}{1+\eta(\tau)}
\label{CtauonEta} \\
 \eta(\tau)&=\frac{C(\tau)}{C_{\Sigma}-C(\tau)}
 \label{CratioFromCtau}
\end{align}
The expression (\ref{CratioFromCtau}) is  mathematically identical to (\ref{Cratio})
but it is not sensitive to a presence of charge leaks in experimental measurements
if the charge $Q$ in $C(\tau)$ from (\ref{Ctau})
is calculated using numerical integration (\ref{Qdc}).

The two--capacitors inverse relaxation model
provides a single exponent behavior of $U(t)$
in the open circuit regime
with $R_2C_1 C_2/(C_1+C_2)$ exponent time.
In a system with a number of porous branches the
behavior of $U(t)$ is more complex.
In the Fig. \ref{threeRC} a three--$RC$ supercapacitor model
is presented. The system has multi--exponents in $U(t)$ evolution,
stable $\eta$ for $\tau\ll RC$ in two orders or magnitude range,
and $C(\tau)$ asymptotic
 $C(\tau=0)=C_1$  and $C(\tau=\infty)=C_1+C_2+C_3+\dots$.

The $C(\tau)$ shows 
what the capacitance the system responds at time--scale $\tau$.
It is measured from $U(t)$ sampling with subsequent
integration (\ref{Qdc}). An important advantage of $C(\tau)$ is
that it requires only $\tau$--long $U(t)$ measurement:  during shorting period
and the potential immediately after switching to open circuit regime,
this makes (\ref{Ctau}) non--susceptible
to charge leaks (no $U_2$ measurement is required).
The $C(\tau)$ has a very clear practical meaning:
If partial discharge takes time $\tau$,
what is the ratio of charge/potential change for the $\tau$.
This makes the inverse relaxation technique a well suitable tool
for characterization
of supercapacitors 
and other relaxation type systems with porous structure.

\section{\label{sc}The Experimental Measurement Of Supercapacitors}
The circuit in Fig. \ref{Utshorting} provides
an implementation of the inverse relaxation measurement technique.
It consists of two computer--controlled switches ``Charge'' and ``Short''
(they can be either MOSFET transistors or fast mechanical relays);
the output potential $U(t)$ is measured by ADC port of a computer.
If a controller
has more than a single ADC then it is convenient
also to record the potential $U^*(t)$ directly from $R_s$, this allows
to
increase measurement precision
and we can avoid (\ref{estimationRTs}) calibration  of the $R_s$, which
is required when a single potential $U(t)$ is recorded,
in which case
the resistance of ``Short'' switch is combined
with the $R_s$.

Measured potentials, before connecting to ADC ports,
must be passed through an operational amplifier with MOSFET
input, e.g. \href{https://www.analog.com/media/en/technical-documentation/data-sheets/AD823.pdf}{AD823},
to decrease parasitic discharge and, especially
for two--potential measurement setup, we can set operational amplifier
to a constant amplification
to bring small potential on shorting stage to
the range of maximal ADC precision.
This setup allows to overcome most of the difficulties\cite{allagui2018short}
of charge measurement at low $U$.
For example, the $R_s$ can be chosen such a small
value that maximal value of $U^*(t)$ Fig. \ref{RCscheme},
will be about $10mV$.
An operational amplifier
\href{https://www.analog.com/media/en/technical-documentation/data-sheets/AD823.pdf}{AD823}
can be used to bring it to a standard ADC max level $3.3V$
with the gain set to just $330$  (a very stable regime) and
$U(t)$ numerical integration (\ref{Qdc}) then gives an accurate
 estimation of the total charge passed.
Formally, the inverse relaxation potential $U_2$
gives the total charge as in Eq. (\ref{CtauSum}).
While one might think this allows
to avoid a
microcontroller--implemented
numerical integration (\ref{Qdc}),
our experimental
measurements show that numerical integration gives
a much more accurate total charge
since it is not sensitive to charge leaks.

When working in a setup of single--potential recording the
``on''--resistance of the ``Short'' switch is combined with the $R_s$.
We can consider some  ``effective''
$R_s$ to enter (\ref{Ri}) and the ``Short'' switch to be ideal.
To obtain the value of an ``effective'' $R_s$ correctly
one can either:
\begin{itemize}
\item Do a calibration
to total charge:
\begin{align}
R_s&\approx \frac{1}{C U_0}\int\limits_0^{\tau} U(t) dt & \tau \gg RC
\label{estimationRTs}
\end{align}
\item Disconnect SC, put ``Short'' and ``Charge'' switches to ``on'' state,
and connect the $U_0$ terminal to a fixed current source, typically about a few ampere. Measured potential $U$ determines the value of effective $R_s$,
this is a variant of
four--terminal sensing technique.
\end{itemize}
In the setup used by the authors an ``effective'' $R_s$ was $0.020\Omega$.

\begin{figure}[b]
\includegraphics[width=12cm]{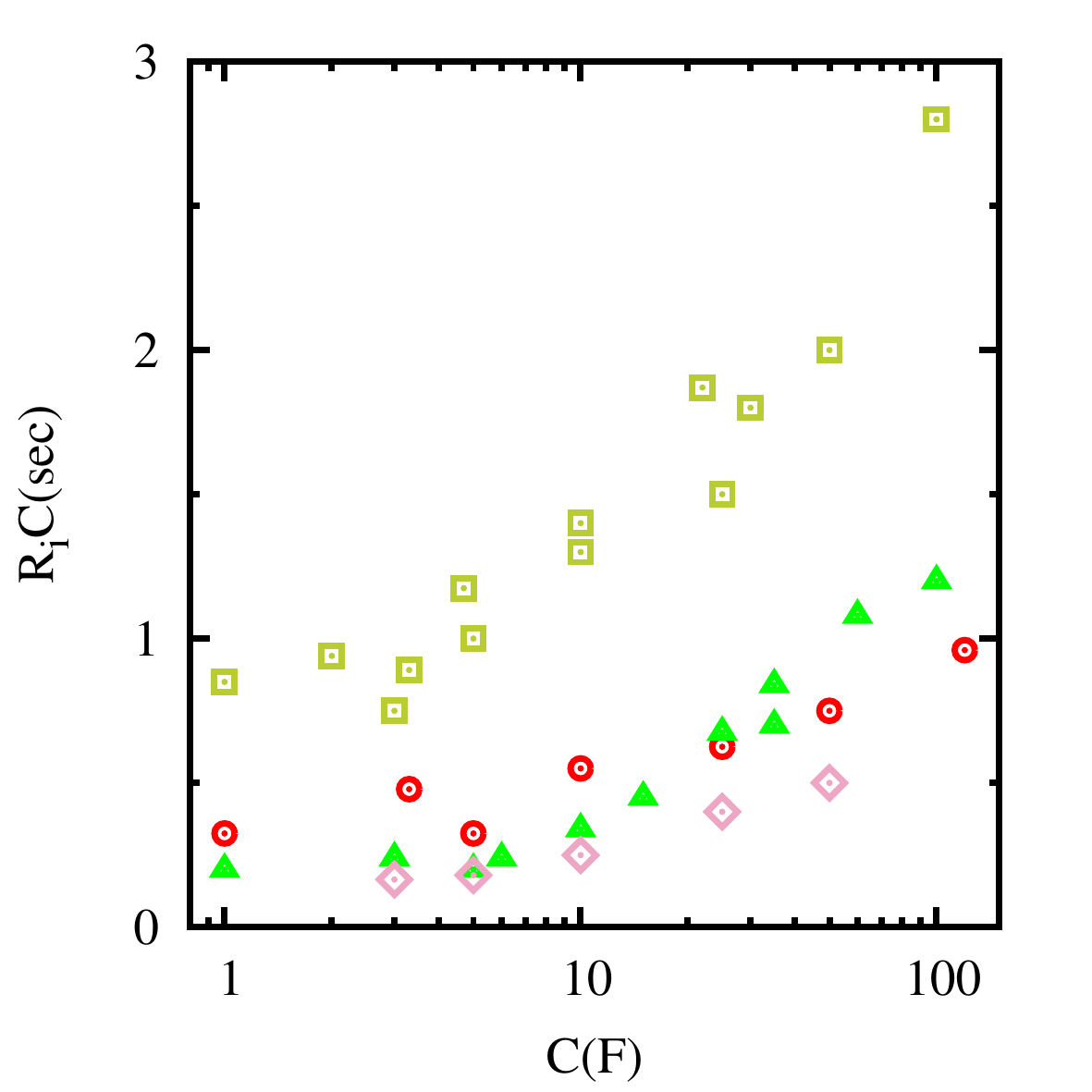}
  \caption{\label{rcdatasheet} The  internal $R_iC$ time as
  a function of capacitance according to datasheet
    for commercial supercapacitors of the manufacturers:
     \href{http://datasheets.avx.com/AVX-SCC-LE.pdf}{AVX}
     (circles),
     \href{https://www.eaton.com/content/dam/eaton/products/electronic-components/resources/data-sheet/eaton-hv-supercapacitors-cylindrical-cells-data-sheet.pdf}{EATON} (triangles),
     \href{https://ru.mouser.com/datasheet/2/88/DCN_series-553005.pdf}{IC} (squares),
     and \href{https://www.maxwell.com/images/documents/ProductMatrix.pdf}{Maxwell} (rhombuses).
     The $R_i$ is taken from the fields  ``ESR Max \@DC'',
     ``Maximum initial ESR'',  ``DC ESR'', and ``ESR DC Typical (10ms)'' respectively.
  }
\end{figure}

\begin{figure}[h]
  \includegraphics[width=12cm]{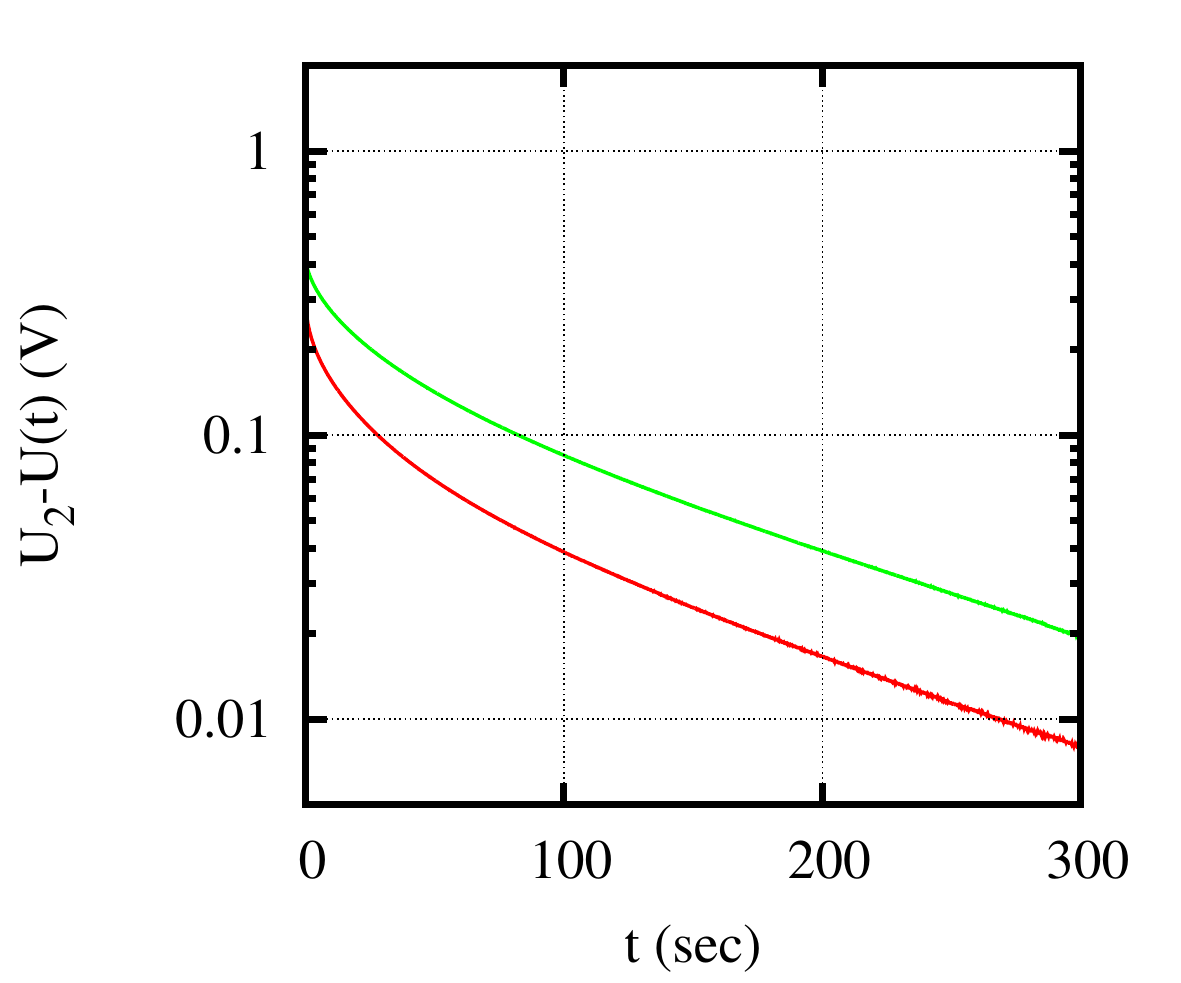}
   \caption{\label{experimentatime} The $U_2-U(t)$ (in $\log$ scale)
    for \CapBlueLink{} (red) and \CapGreenLink{} (green). A deviation from a linear dependence is clearly observed.
    A ``noise'' observed
    at large $t$ is due to exponentially small value of $U_2-U(t)$,
    the situation can be improved by using a high quality ADC.
  }
\end{figure}

We tested  four commercial supercapacitors:
    \CapBlueLink{},
    \CapGreenLink{},
    \CapBlackLink{},
    and \CapWhiteLink{};
all $5F$ with $2.7V$ max.
In Fig. \ref{rcdatasheet}  internal $R_iC$ time
is presented as a function of nominal capacitance
for supercapacitors of the same series 
according to
manufacturer datasheets.
 The $R_iC$ depends mostly on the technology used and increases slowly with the capacitance.
 Supercapacitors have a developed internal structure,
 which manifests itself in multi--exponent $U(t)$ dependence on inverse relaxation stage,
 see Fig. \ref{experimentatime}, which
illustrates
 that inverse relaxation is typically \textbf{not} a
single--exponent type of behavior.
The relaxation at small time is faster
than at large time. An ultimate situation of such a behavior is presented
in Fig. \ref{threeRC}a for a model system.
The deviation $\log(U_2-U(t))$
from a linear law
is related to a distributed internal $RC$.
Fitting of a multi--exponent $U(t)$ relaxation
is a common field of study\cite{il2020modeling}.
A deviation from a single exponent (linear dependence in $\log$ axis)
can be used as a source of information
about supercapacitor's internal structure.
However, such an approach is more susceptible to measurement errors\footnote{
There is a much more advanced Radon--Nikodym technique\cite{2016arXiv161107386V}
that can be applied to obtain relaxation rate distribution as
matrix spectrum
for relaxation type of data such as in 
Fig. \ref{experimentatime}.
The distribution of the eigenvalues
(using the Lebesgue quadrature\cite{ArxivMalyshkinLebesgue}
weight as eigenvalue weight)
is an estimator of the distribution
of relaxation rates observed in the measurement;
Radon--Nikodym approach is much less susceptible to measurement errors
compared to inverse Laplace
transform type of analysis.
See \cite{liionizversiyaran}
for 
application example to Li--Ion degradation rate estimation.
}
and has interpretation difficulty.


\begin{figure}[h]
 \includegraphics[width=10cm]{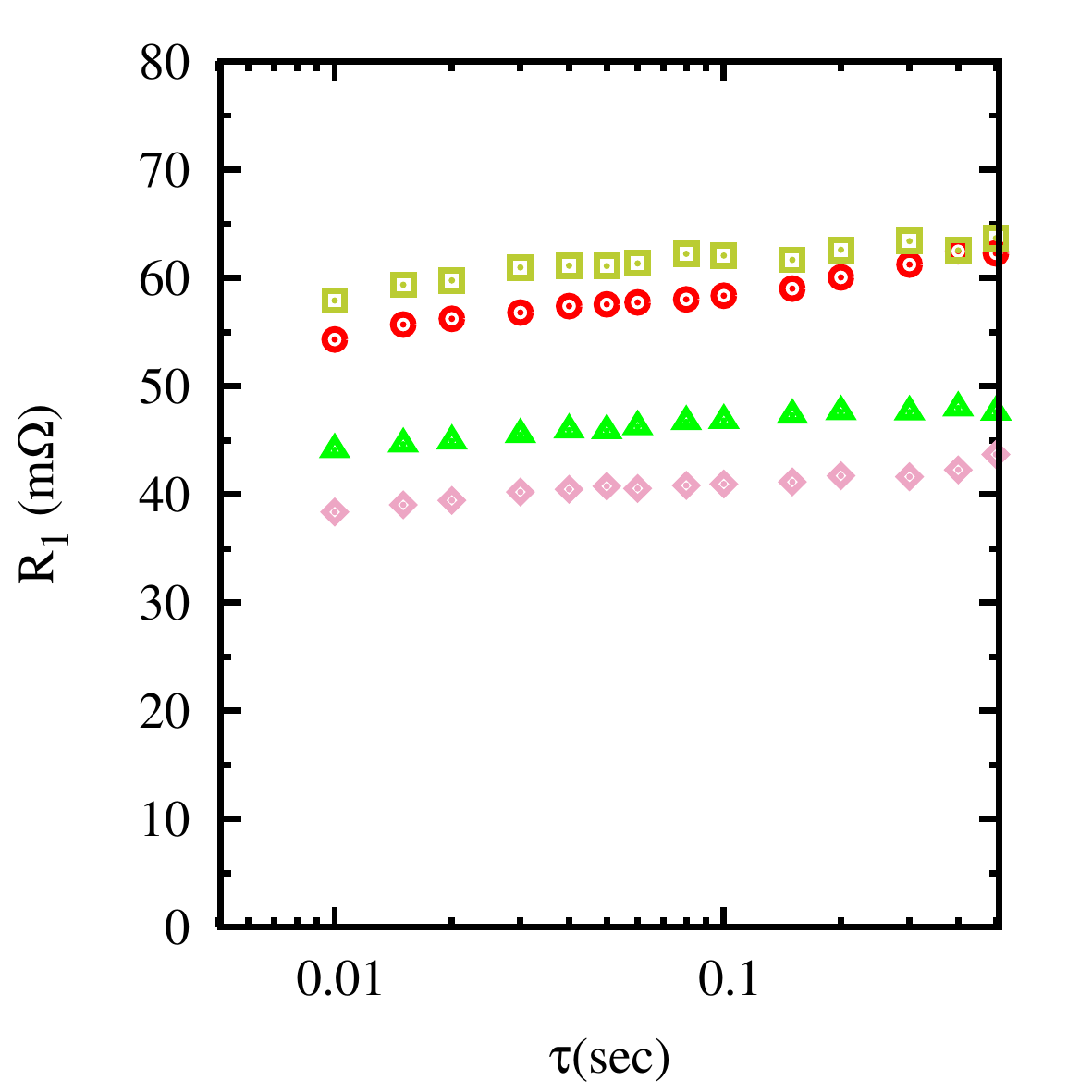}
 \caption{\label{InternalRCQ}
   An internal resistance component $R_1$ from
   (\ref{Ri}) measurement for 
    four commercial supercapacitors
    \CapBlueLink{} (circles),
    \CapGreenLink{} (triangles),
    \CapBlackLink{} (squares),
    and \CapWhiteLink{} (rhombuses);
     as a function on $\tau$.
  }
\end{figure}

Let us start discussing our experimental results
with internal resistance $R_1$, Eq. (\ref{Ri}),
it is presented in Fig. \ref{InternalRCQ}.
Measured $R_1$ does not depend on shorting time at all,
the value corresponds to minimal possible internal resistance.
Some manufacturers present the
internal resistance for different $\tau$ (e.g. \cite{maxwellSC}
presents internal resistance for $0.01$sec and $5$sec).
In their setup (which assumes constant capacitance and variable resistance)
larger $\tau$ corresponds
to current propagation to supercapacitor deep pores,
what involves a contribution from $R_2$.
In our setup we have a constant $R(\tau)=R_1=const$
and instead consider an ``effective'' \textsl{capacitance} as a function of $\tau$,
this is how the $C(\tau)$  is defined in Eq. (\ref{Ctau}).
It has
a very clear meaning: the ratio of charge/potential change for the $\tau$;
this corresponds to a typical SC setup: discharge as much as you can
in a given time $\tau$.

\begin{figure}
  \includegraphics[width=8cm]{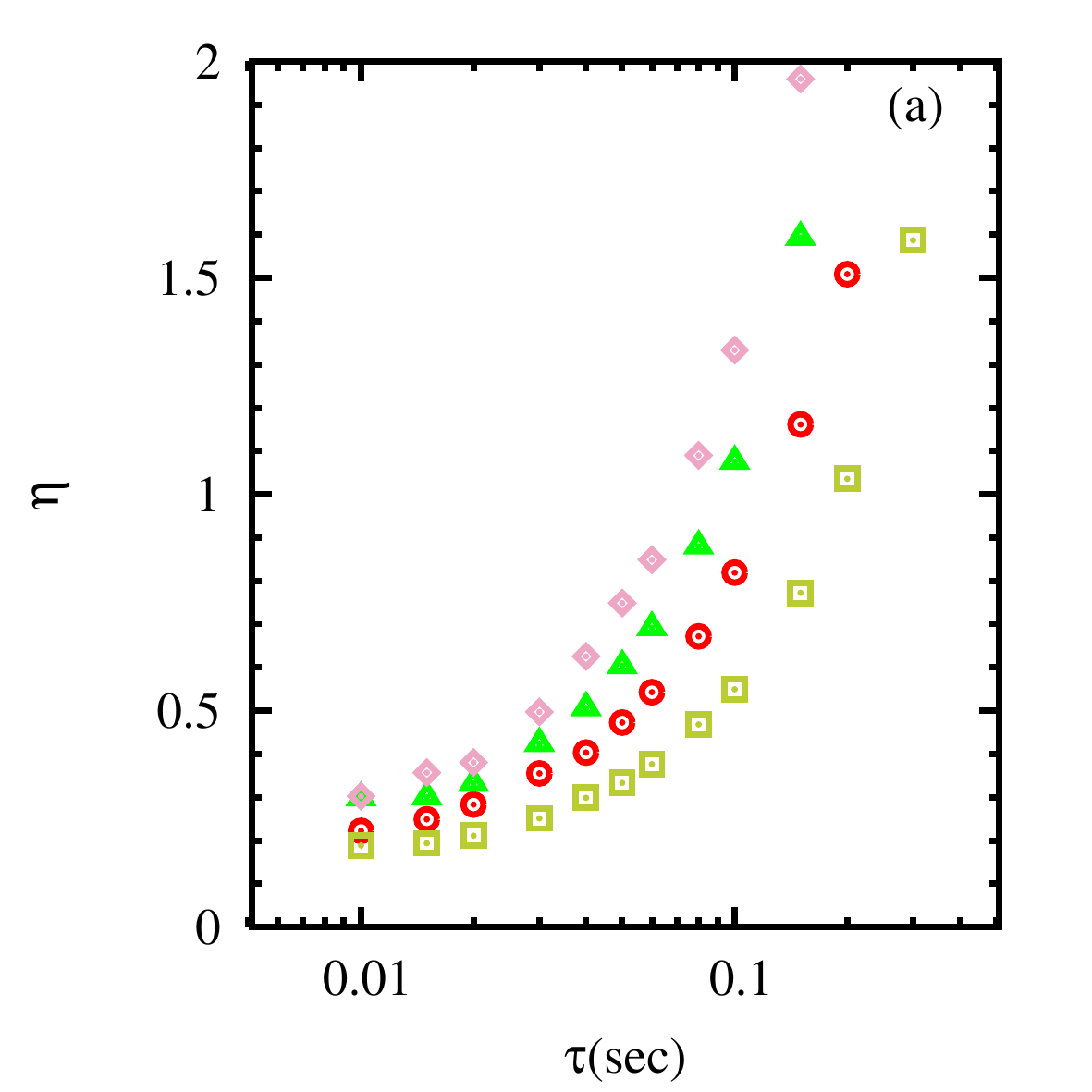}  
  \includegraphics[width=8cm]{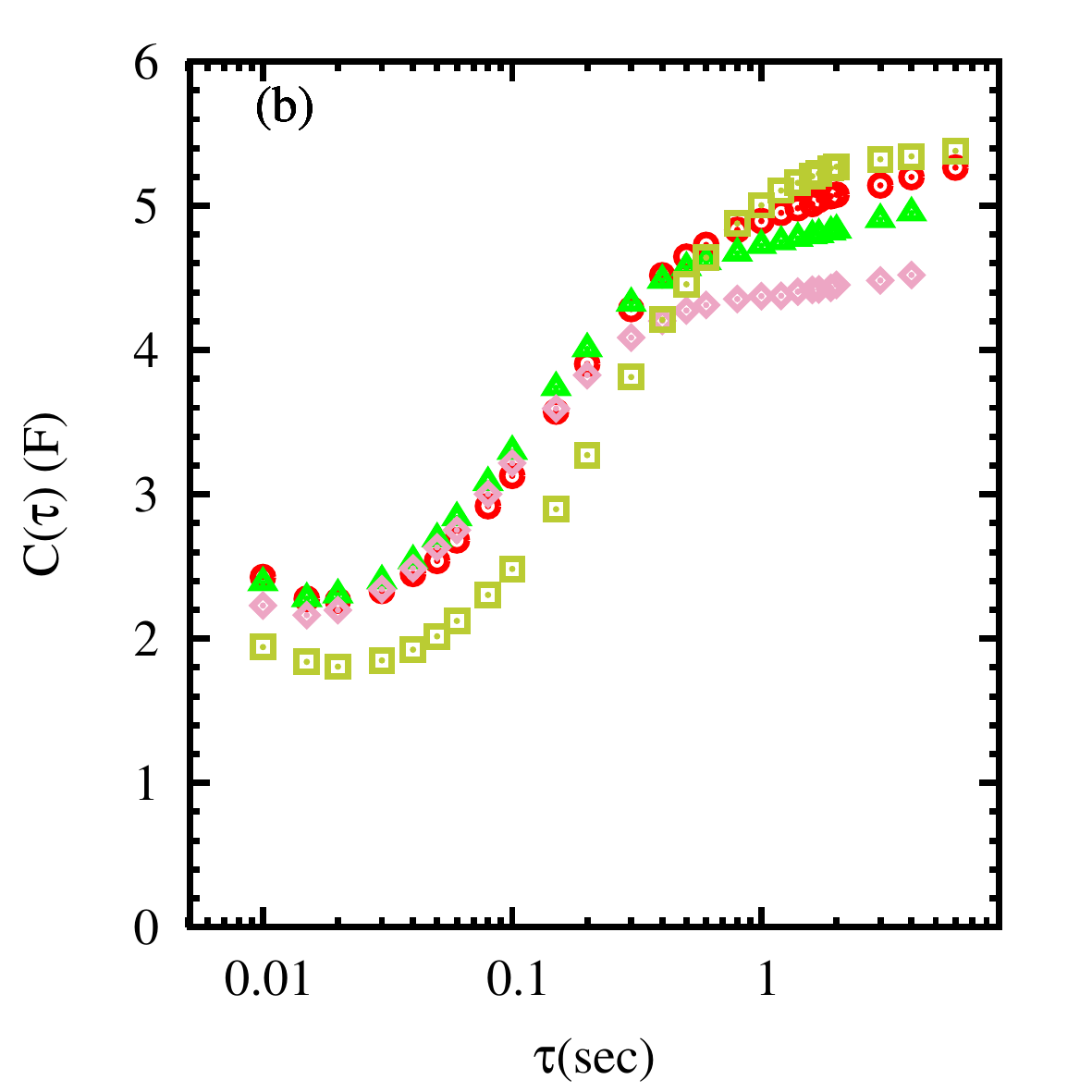}
\caption{\label{etaCcommercial}
Four commercial $5F$, $2.7V$ supercapacitors:
    \CapBlueLink{} (circles),
    \CapGreenLink{} (triangles),
    \CapBlackLink{} (squares),
    and \CapWhiteLink{} (rhombuses).
     The  dependence      of
     (a): $\eta$ (\ref{Cratio}) on shorting time
    $\tau$;
    (b): $C(\tau)$ (\ref{Ctau}) on shorting time
    $\tau$.
 In this chart the $C(\tau)$ is measured much more accurately than the $\eta$
    because (\ref{Ctau}) uses numerical integration (\ref{Qdc})
    and (\ref{Cratio}) uses $U_2$
    to calculate the charge passed during shorting stage.
    The (\ref{CratioFromCtau}) is a more accurate method to obtain $\eta$.
}
\end{figure}

In Fig. \ref{etaCcommercial}a
$\eta$ as a function of shorting time is presented.
Only three potentials $U_0$, $U_1$, and $U_2$
have to be measured; no measurement of
exponentially small values is required:
the $U_0-U_1$ and $U_2-U_1$ are not small for
actual $\tau$ values used in the experiment.
The potentials $U_0$, $U_1$, and asymptotic $U_2$
are measured directly.
The only difficulty that may arise if a supercapacitor
has a parasitic charge leak,
both internal and through $U(t)$ measurement circuit;
this affects $\eta$ (\ref{Cratio}) and $C_{\Sigma}$ (\ref{CtauSum}),
but does not affect the $C(\tau)$ (\ref{Ctau}).
A good heuristic for $U_2$ in case of a substantial self--discharge
is the maximal value of $U(t)$ on inverse relaxation stage.
A maximal $\tau$, for which a plateau can still be observed,
is the value below characteristic scale of inverse relaxation.
For a typical supercapacitor a
characteristic scale of the inverse relaxation
can be estimated
as a \textsl{multiple} of internal $R_iC$ (available from Fig. \ref{rcdatasheet}
at $C=5F$ for the supercapacitors we use).
This scale is different from $R_iC$ (typically several times greater), but
about the same order of magnitude.

In Fig. \ref{etaCcommercial}b measured $C(\tau)$ dependence is presented.
This is the most informative chart of the
inverse relaxation technique.
For all four supercapacitors ($5~{\rm F}$ nominal)
low $\tau$ discharge is equivalent to a discharge
of about $2F$ ideal capacitor, then the $C(\tau)$ increases with $\tau$.
At high $\tau$ the discharge
is equivalent to a $5F$ nominal capacitor.
The $C(\tau)$ at high $\tau$
being equal to nominal capacitance
demonstrates correct operation of ADC and 
numerical integration (\ref{Qdc}).
A selection of insufficiently small $R_s$ may introduce
an error to the value of $\tau$ at which $C(\tau)$ starts to increase.
Typically the internal relaxation resistance is much greater
than the internal resistance (i.e. $R_2\gg R_1$ for a simple two--$RC$ model in Fig. \ref{RCscheme}). For $R_s$ a good choice is lower (or even about) than the internal resistance $R_i$ (for a simple two--$RC$ model it is $R_s\lesssim R_1$). If $R_s$ is chosen  greater than the internal relaxation resistance  ($R_2$ in two--$RC$ model) then the plot $C(\tau)$ will be shifted to larger $\tau$,
thus a ``too high $R_s$'' measurement setup
 will underestimate the qualtity of a supercapacitor.

\subsection{\label{impedance}Impedance characteristics of the supercapacitors}

The biggest advantage
of impedance spectroscopy is that it 
can capture a wide range (many orders) of frequencies\footnote{The biggest advantage
of impedance spectroscopy
is that impedance function
is a ratio of two polynomials,
thus it can be measured/interpolated/approximated with a
high degree of accuracy for
the measurements in a wide range (over $9$ orders, typically $10^{-3}\div10^{6}$Hz) of frequency responses.
However, in time--domain, where exponentially small values need to
be measured, a much smaller range of time--scales are accessible (less than $2$ orders, often just a single order), 
hence in standard mathematical techniques, such as inverse Laplace transform,
any type of noise/discretization/measurement error/window effect
have a huge impact on 
exponentially small Laplace transform contributions\cite{2016arXiv161107386V}}.
The disadvantages of the technique
are: measurement equipment complexity, impedance interpretation
difficulty,
and typically a low current linear regime, thus
non--linear effects are problematic to study\cite{kompan2010nonlinear}.
Most manufacturers provide equivalent ESR at fixed frequency $1000Hz$
in the datasheets,
which is typically several times lower than the one at DC. In this section
we apply impedance technique to obtain DC characteristics
of supercapacitors. The goal
is to compare impedance approach with inverse relaxation.

\begin{figure}[ht]
\includegraphics[width=8cm]{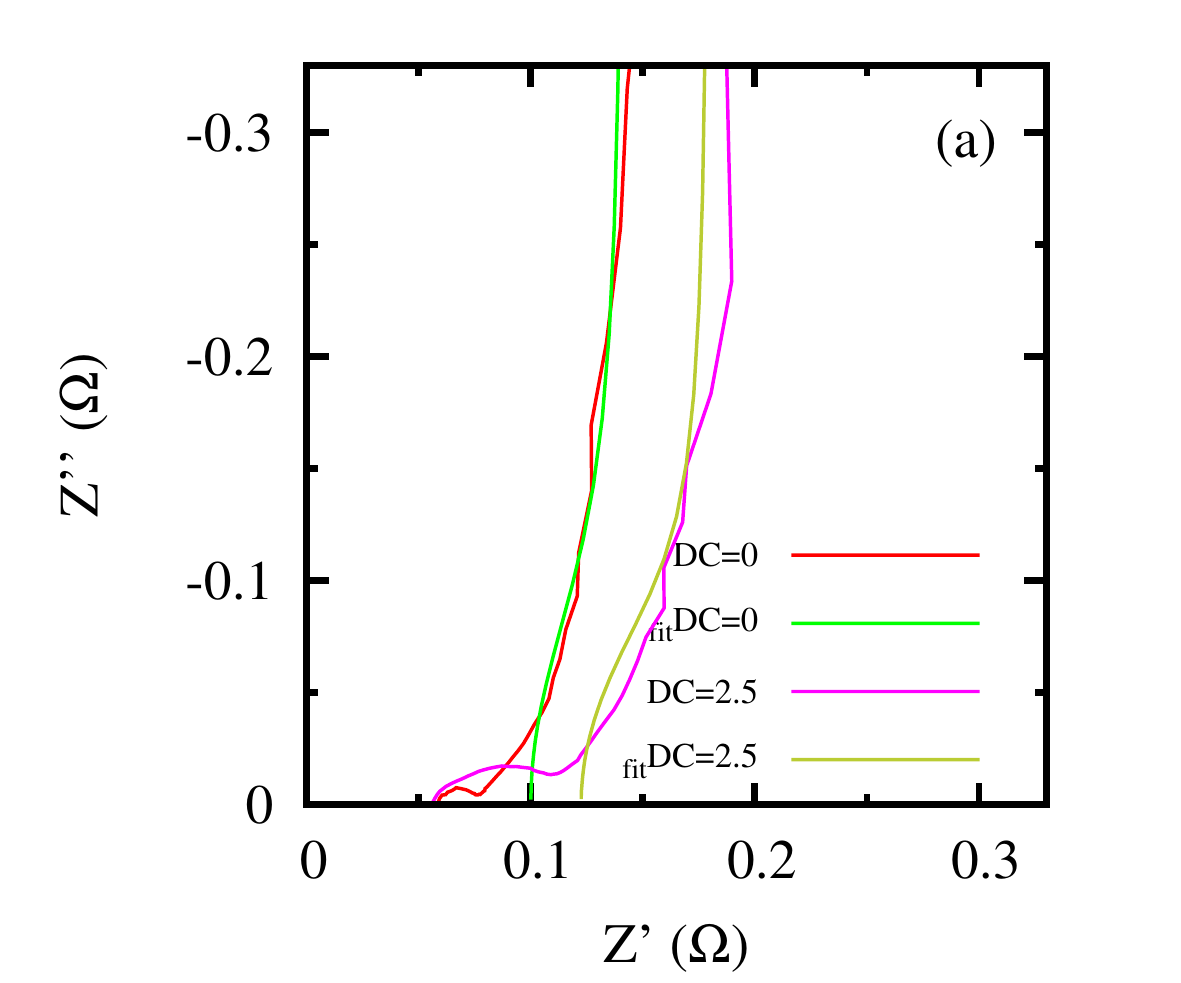}
\includegraphics[width=8cm]{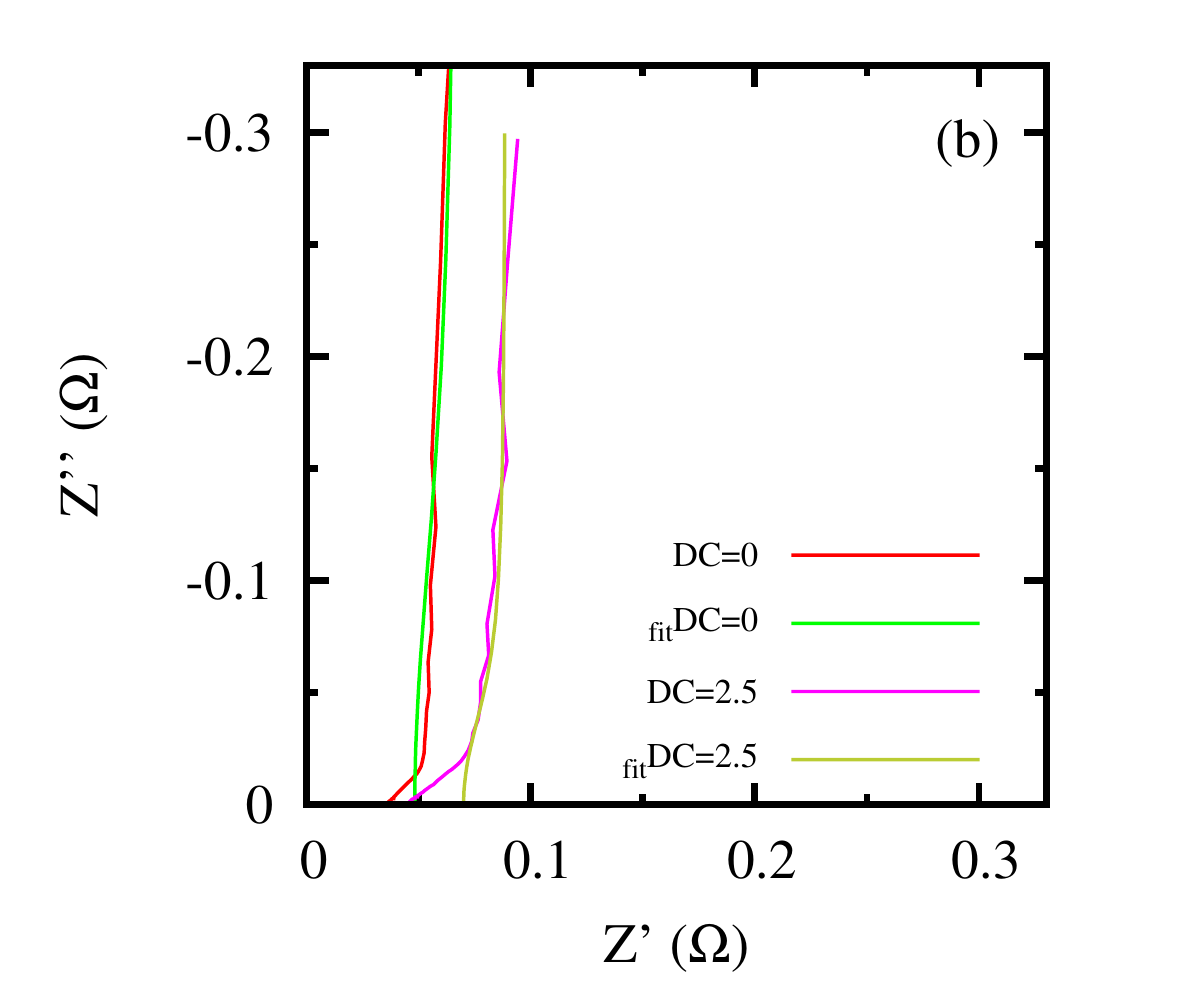}
  \caption{\label{FigImpedance}
    Impedance curves for two
    commercial supercapacitors
    with potential offsets $DC=0V$ and $DC=2.5V$.
    The curves are then fitted with two chain $RC$ model in Fig. \ref{RCscheme}
using \href{https://www.ameteksi.com/products/software/zview-software}{ZView}
program,
the values ($R_1$, $C_1$, $R_2$, $C_2$) are obtained from the fitting.
Frequency range has been chosen
as $10^{-3}\div 30$Hz;
the impedance was measured at very small $5mV$ AC amplitude.
(a):  \CapBlueLink{} $DC=0V$: $(0.1\Omega, 3.765F, 0.3961\Omega, 1.334F)$ $\eta=2.82$,
      $DC=2.5V$:$(0.122\Omega, 2.5F, 0.24\Omega, 2.35F)$ $\eta=1.06$.
(b): \CapGreenLink{}:
    $DC=0V$: $(0.048\Omega, 3.797F, 1.527\Omega, 0.489F)$ $\eta=7.76$,
      $DC=2.5V$: $(0.065\Omega, 3.79F, 0.3\Omega, 1.737F)$ $\eta=1.95$.      
}
\end{figure}

A common impedance analysis method
is \href{https://en.wikipedia.org/wiki/Nyquist_stability_criterion#Nyquist_plot}{Nyquist plot}.
In Fig. \ref{FigImpedance}
the Nyquist plot $Z^{\prime}Z^{\prime\prime}$
 is presented along with ZView fitting by
 two--$RC$ model
 for  \CapBlueLink{} and  \CapGreenLink{}
 supercapacitors.
The impedance measurements have been performed in a frequency
 range $10^{-3}\div 10^{5}$Hz.
 In this range
Nyquist plot has a complex behavior caused by a complex internal structure of the device.
In supercapacitor applications
the frequencies of practical interest
are the ones below $30\div 50$Hz.
For simple models (such as in Fig. \ref{RCscheme})
 it would be rather na\"{\i}ve trying to
fit many orders of frequency range by a simple circuit of several $RC$ chains.
For these reasons we limit the frequency range
by $10^{-3}\div 30$Hz. A simple one--$RC$ model
has a vertical asymptotic behavior at low frequencies.
Two--$RC$ chains give some slope at low frequencies,
observed in Fig. \ref{FigImpedance}.
In ZView, a model with two PCE elements (one with small exponent,
a second one
is close to $1$, almost the capacitance)
allows to obtain a very good fit
of the impedance curve  in the entire $10^{-3}\div 10^{5}$Hz frequency range.
The PCE element by itself
can be modeled  as a sequence of $RC$ elements\cite{valsa2013rc},
thus the value and exponent of PCE describe the supercapacitor's
internal structure. However, a limited range of
practically interesting frequencies along with
interpretation difficulties makes this approach not very appealing.

A very important feature of a supercapacitor,
not observed in a regular capacitor,
is that the impedance curve
depends on DC potential applied.
When  the DC potential changes from $0V$ to $2.5V$
the impedance curve shifts to the right (the supercapacitor's
internal resistance increases) and
the Nyquist plot changes substantially.
The dependence of the capacitance on the applied potential
is a known effect. It can be caused  by the density of state changes\cite{kompan2015ultimate},
 double layer structure changes\cite{kornyshev2007double,bagotsky2015electrochemical,zhan2017computational,bossa2018differential},
or redox--active electrolyte processes\cite{dai2016voltage,ban2013charging}
 of both reversible (Faraday's capacitance) and
  irreversible (electrochemical decomposition) types.

The data, obtained from  $10^{-3}\div 30$Hz impedance
fitting differ quite substantially from the results
of previous section.
While the value of $R_1$ and total capacitance are similar, the 
 $\eta$ differ substantially, also it typically decreases
 with DC potential increase:
changes
from $2.82$ to $1.06$ for \CapBlueLink{}
and from $7.76$ to $1.95$ for \CapGreenLink{},
the same behavior was observed in the other supercapacitors we measured.

These measurements make us to conclude
that an approach of ``stretching'' small signal impedance technique down to DC
range is not a  good idea.
The inverse relaxation has important advantages of being
close to ``natural'' fast discharge
regime of a supercapacitor deployment and the measurement technique itself
is much simpler than the impedance technique.

\section{\label{Discussion}Discussion}

In this work a novel technique for supercapacitors
characterization is developed, modeled numerically,
and experimentally tested on a number of commercial supercapacitors.
The technique does not have any exponentially small value to measure,
while, in the same time, all the measurements
are performed not in the frequency domain,
but in the time domain; the measurement is technically feasible
in at least three orders of time--scales: $10^{-2}\div 10^1$sec.
Besides of the simplicity of the technique (no fitting is required),
the most important feature
of the inverse relaxation approach
is it's simple automation. Microcontroller--operated two switches and
a single ADC can obtain the dependence of
an ``effective'' capacitance on time--scale $C(\tau)$,
 Fig. \ref{etaCcommercial}b. 
The approach can be considered as an alternative to
commonly used\cite{maxwellSC}
consideration of a constant capacitance and time--scale
dependent internal resistance. Among the advantages of our technique
is that it does not require a source of fixed current,
what simplify the setup and allows a very high discharge current
regime. As the limitations of the techniques we would note:
only about three order of available $\tau$ range
(it is difficult to measure at shorting time $\tau\lesssim 10ms$)
compared with about nine orders $10^{-3} \div 10^6$Hz of available
frequency range in impedance spectroscopy,
the necessity to use a microcontroller for numerical integration (\ref{Qdc}),
and a risk to destroy a supercapacitor on shorting stage
by a high discharge current.

Modeling supercapacitors internal structure
in electronic circuit software
is a common field of study\cite{johansson2008comparison,logerais2015modeling,pean2016multi,song2019equivalent,il2020modeling}.
In \cite{fletcher2017modelling} a voltage rebound
effect, shorting and then
switching to open circuit was also modeled.
However, only in our early work\cite{kompan2019reverse} the
ratio $\eta(\tau\to 0)$
of ``easy''  and ``hard'' to access capacitance
was introduced.
Similar pulse--response characteristics of Li--ion batteries
have been studied in \cite{barai2018study}
with an emphasis on time--scale.

The idea to use pulsed load for a battery (primary or secondary) or fuel cell
is now widespread. For example
GSM standard specifies $0.575ms$ transmission burst
within a $4.6ms$ period (1/8 duty factor),
thus DC--DC chips like \cite{maximDCDCGSM},
that are especially designed for pulsed load,
have been used in all modern devices.
S.L.Kulakov pioneered an application of pulsed load
to metal--air power sources and then, about a decade later,
brought to our attention the pulsed load technique in \cite{danielyan2007increasing}.
Developed in this paper pulsed technique for supercapacitors
characterization is dedicated to his memory.

\appendix

\section{\label{psiX}Software Modeling}
The systems have been modeled in
\href{http://ngspice.sourceforge.net/}{Ngspice circuit simulator}.
The circuit have been created in  gschem  program of
\href{http://www.geda-project.org/}{gEDA} project.
To run the simulator download\cite{RCsimulator} the file
\href{http://www.ioffe.ru/LNEPS/malyshkin/RCcircuit.zip}{RCcircuit.zip}
and decompress it.
To test the simulator execute
\begin{verbatim}
ngspice  Farades_y_with_variables.sch.autogen.net.cir
\end{verbatim}
Because original gschem+ngspice
do not have a convenient parameterisation, a perl script \verb+run_auto.pl+
have been developed.
To run the simulator with $\tau=0.2$sec execute:
\begin{verbatim}
perl -w run_auto.pl Farades_y_with_variables.sch   0.2
\end{verbatim}
The script takes the \verb+Farades_y_with_variables.sch+
which corresponds to three--$RC$ system in Fig. \ref{threeRC}
and substitute shorting time  $\tau=0.2$.
One can modify it accordingly, and run ngspice.
The result is saved to \verb+n0_output.txt+.

\bibliography{echem,LD}

\end{document}